**Viral replication dynamics could critically modulate vaccine effectiveness and should be accounted for when assessing new SARS-CoV-2 variants**

Wan Yang[1*] and Jeffrey Shaman[2]
[1]Department of Epidemiology, [2]Department of Environmental Health Sciences, Mailman School of Public Health, Columbia University, New York, NY, USA
*Correspondence to: wy2202@cumc.columbia.edu

**Abstract:**
In this article, we propose a theory to explain the reduction in vaccine effectiveness (VE) against the Delta SARS-CoV-2 variant and decreasing VE over time reported in recent studies. Using a model illustration, we show that in-host viral replication dynamics and delays in immune response could play a key role in VE. Given this, current laboratory approaches solely measuring reductions in neutralizing ability cannot fully represent the potential impact of new SARS-CoV-2 variants. We instead propose an alternative approach that incorporates viral replication dynamics into evaluations of SARS-CoV-2 variant impact on immunity and VE. This more robust assessment may better inform public health response to new variants like the newly detected Omicron variant.

**Text:**
Recent studies have reported reductions in COVID-19 mRNA vaccine effectiveness (VE) against infection and, to a lesser extent, against severe disease for the Delta SARS-CoV-2 variant of concern (VOC). These reductions in VE – particularly against infection – have been attributed to waning immunity, which appears to be supported by two lines of evidence: 1) laboratory studies comparing neutralizing ability of vaccinee sera against different variants; and 2) estimated VE against Delta that decreases as the time since vaccination increases.[1,2] Notably, the reduction of vaccinee sera neutralizing ability against Delta is smaller than Beta (2-8 fold reduction against Delta vs. 10-40 fold against Beta),[3,4] yet more breakthrough infections have been observed for Delta,[5] coincidental with its later emergence and circulation. These findings have prompted administration of a 3rd (i.e. booster) vaccine dose, and preliminary results have shown restored VE shortly after this boosting.[6] However, it remains unclear why the substantial reduction in VE occurred mostly when Delta became the predominant circulating variant and why VE against severe disease is more preserved than infection.

We hypothesize that in-host viral replication dynamics and delays in immune response play a key role in VE. Compared to ancestral variants, Delta not only carries mutations enabling some degree of immune escape but also faster replication that leads to a much higher viral load in infected individuals (e.g., 10 to 1000 times higher than ancestral variants[7,8]). As such, by the time adaptive immunity ramps up in response to infection, generated antibody titers need to be 2 to 3 orders of magnitude higher in order to neutralize the abundance of Delta virions. For example, if we combine neutralization reduction and faster replication, a 4-fold reduction in antibody affinity and a 100-fold increase in viral load requires antibody titers 400 times higher, as opposed to 4 times higher as suggested by antibody neutralizing experiments alone. If Delta replication initially outpaces elevation of antibody titers, symptomatic infection may be more likely; however, as adaptive immunity continues to ramp up, it will ultimately overpower Delta and prevent more severe disease outcomes. This hypothesis would explain the loss of protection against symptomatic infection by Delta over time, but the resilience of protection against severe, critical and fatal disease.[2]

To illustrate this hypothesis, we simulated in-host viral dynamics for a hypothetical wildtype infection and a Delta-like VOC infection among unvaccinated, not recently vaccinated, and recently vaccinated or boosted individuals. As shown in Fig 1, the Delta-like VOC generates reduced but still substantially high viral loads in non-recent vaccinees (vs. naïve / unvaccinated) due to its faster replication rate and the slight delay of immune response. Further, viral load is substantially reduced in recently vaccinated or boosted individuals due to higher circulating titers and a faster and stronger immune response (Fig 1). These simulated results are consistent with the observed higher probability of Delta breakthrough but lower probability of severe infection, as well as the restored VE following the boosting.

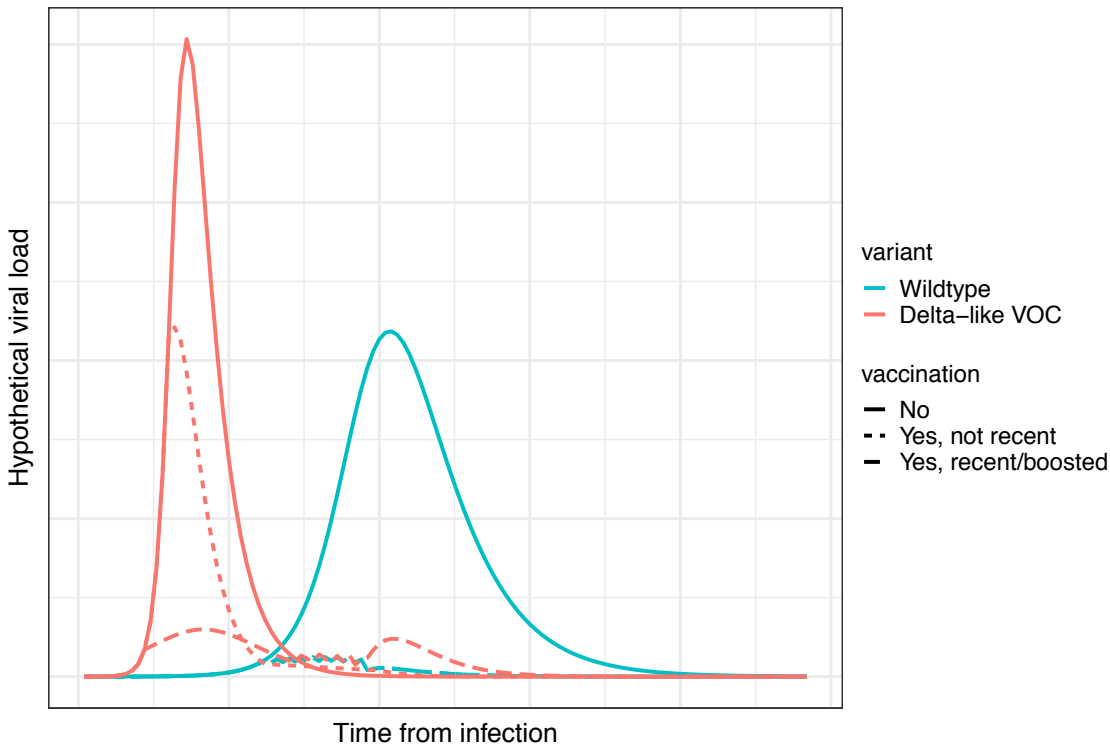

**Fig 1. Simulated in-host viral dynamics.** The simulation assumes 1) a basic reproduction number $R_0$ of 2.4 and an infectious period $D$ of 6 days for the wildtype infection and an $R_0$ of 4 and $D$ of 4 days for a Delta-like VOC infection; 2) lower antibody affinity for the Delta-like VOC leading to a 15% reduction in immune response; and 3) for the recently vaccinated or boosted, immune response is 25% faster and stronger than those not recently vaccinated.

Given the above, we call for more robust assessment of SARS-CoV-2 variant impact on immunity and VE. Specifically, viral replication dynamics need to be incorporated when assessing sensitivity to antibody neutralization. For instance, in addition to assays using the same starting viral titer, neutralizing experiments can be performed using harvested virus from cell cultures inoculated with the same amount of virus and grown for a certain period of time (e.g. roughly the time from infection to immune response). While such a study design also has limitations (e.g., *in vitro* cell culture may not fully represent *in vivo* replication dynamics), it may better represent the combined outcome of in-host viral dynamics and delayed immune response. Such findings can better inform public health response to new variants like the newly detected Omicron variant.